\documentclass[11pt,3p,review,authoryear]{elsarticle}

\makeatletter
\def\ps@pprintTitle{%
  \let\@oddhead\@empty
  \let\@evenhead\@empty
  \let\@oddfoot\@empty
  \let\@evenfoot\@oddfoot
}
\makeatother

\usepackage{amsmath}
\usepackage{graphicx,psfrag,epsf}
\usepackage{enumerate}
\usepackage{float}
\usepackage[flushleft]{threeparttable}
\usepackage[hidelinks]{hyperref}
\usepackage{xcolor}

\journal{CES 2024}
\biboptions{longnamesfirst}
\begin{document}

\begin{frontmatter}

\title{Free to play: UN Trade and Development's experience with developing its own open-source RAG LLM application}

\author[ss]{Daniel Hopp\corref{cor}}
\address[ss]{Data Scientist\\
UN Trade and Development\\
Palais des Nations ONU, Office E 10035\\
Avenue de la Paix 8, CH-1211 Genève 10\\
Prepared for the seventy-second plenary session of the Conference of European Statisticians}

%% Only give the email address of the corresponding author
\cortext[cor]{Daniel Hopp}
\ead{daniel.hopp@un.org}

\begin{abstract}
Generative AI, and in particular large language models (LLMs), have exploded in popularity and attention since the release to the public of ChatGPT's GPT-3.5 model in November of 2022. Due to the power of these general purpose models and their ability to communicate in natural language, they can be useful in a range of domains, including the work of official statistics and international organizations. For instance, they can help users generate queries on statistical databases to interact with official statistics in a more natural, user-friendly manner. However, with such a novel and seemingly complex technology, it can feel as if generative AI is something that happens to an organization, something that can be talked about but not understood, that can be commented on but not contributed to. Additionally, the costs of adoption and operation of proprietary solutions can be both uncertain and high, a barrier for often cost-constrained international organizations. In the face of these challenges, UN Trade and Development (UNCTAD), through its Global Crisis Response Group (GCRG), has explored and developed its own open-source Retrieval Augmented Generation (RAG) LLM application. RAG makes LLMs aware of and more useful for the organization's domain and work. Developing in-house solutions comes with pros and cons, with pros including cost, flexibility, and fostering institutional knowledge. Cons include time and skill investments and gaps and application polish and power. The three libraries developed to produce the app, \textcolor{blue}{\href{https://pypi.org/project/nlp-pipeline/}{nlp\_pipeline}} for document processing and statistical analysis, \textcolor{blue}{\href{https://pypi.org/project/local-rag-llm/}{local\_rag\_llm}} for running a local RAG LLM, and \textcolor{blue}{\href{https://github.com/dhopp1/streamlit_rag}{streamlit\_rag}} for the user interface, are publicly available on PyPI and GitHub with Dockerfiles. A fourth library, \textcolor{blue}{\href{https://github.com/dhopp1/local_llm_finetune}{local\_llm\_finetune}}, is also available for fine-tuning existing LLMs which can then be used in the application.
\end{abstract}

\begin{keyword}
AI \sep LLM \sep RAG \sep official statistics
\end{keyword}

\end{frontmatter}

\newpage

\section{Introduction}
\label{sec:introduction}

Ever since ChatGPT's release of its GPT-3.5 model in November, 2022, \textcolor{blue}{\href{https://www.forbes.com/sites/bernardmarr/2023/05/19/a-short-history-of-chatgpt-how-we-got-to-where-we-are-today/}{the public has been captivated by the promise of generative AI}}. Many people are by now familiar with the basic premise of a large language model (LLM), but briefly, they are artificial neural networks (ANNs) trained on huge amounts of natural language text scraped from the internet or other sources. They are then able to predict the next word or series of words based on prior context. The end results are remarkably convincing and reasonable responses to prompts and queries. This combination of a vast knowledge base and flexible natural language input and output have found offerings like ChatGPT a fast home in peoples' bookmarks bars.

LLMs' utility and relevance can be further enhanced by the use of Retrieval Augmented Generation (RAG), which will be discussed in more detail in section \ref{sec:rag}. In short, RAG offers a way for LLMs to be aware of and contextualized on new (in terms of either domain or time) information. This can provide many benefits for organizations in general, as well as for national statistical offices (NSOs). A non-exhaustive list of exemplary use cases is presented below.

\begin{itemize}
    \item Querying against a collection of trusted research on a topic to synthesize the information and have it presented at the register of a desired knowledge level, e.g., getting non-technical summaries and explanations to a technical topic.
    \item Querying against up-to-date code documentation to help with coding for less popular programming languages or libraries, which may be particularly relevant for NSOs, who may use specific software.
    \item Querying against meeting notes, field reports, or other confidential administrative documents to draft summaries. In cases where an official summary is required, drafters could use these initial inputs to speed up their work. In cases where one is not required, LLM-generated outputs could be used directly to summarize work and proceedings that would normally not get such treatment and make them available to a wider audience.
    \item Querying against your own organization's reports to generate insights and drafting help in the style of your organization, as well as retrieve relevant passages and ensure institutional consistency on certain topics in a much more flexible and robust manner than simple keyword search.
    \item Querying against the API documentation and table definitions for your statistical database to allow users easier access to the data. For instance, users could ask questions in natural language like, "I would like the GDP of countries in southern Africa in constant prices from 2012-2020", and have a database query generated, explained, and run automatically.
\end{itemize}

Having established the value of having an LLM that is aware of your own organization's specific body of work or domain, the question becomes how to implement and utilize this. The current landscape of LLM services is split between offerings from large tech companies, such as Google's \textcolor{blue}{\href{https://gemini.google.com/}{Gemini}} or Microsoft's \textcolor{blue}{\href{https://copilot.microsoft.com/}{Copilot}} and \textcolor{blue}{\href{https://www.bing.com/}{Bing}}, or those from smaller companies heavily invested in by larger tech companies, such as \textcolor{blue}{\href{https://chat.openai.com/}{ChatGPT}} (\textcolor{blue}{\href{https://www.forbes.com/sites/qai/2023/01/27/microsoft-confirms-its-10-billion-investment-into-chatgpt-changing-how-microsoft-competes-with-google-apple-and-other-tech-giants/}{Microsoft}}), \textcolor{blue}{\href{https://mistral.ai/}{Mistral}} (\textcolor{blue}{\href{https://www.theverge.com/2024/2/26/24083510/microsoft-mistral-partnership-deal-azure-ait}{Microsoft}}), or \textcolor{blue}{\href{https://www.anthropic.com/claude}{Claude}} (\textcolor{blue}{\href{https://www.cnbc.com/2023/10/27/google-commits-to-invest-2-billion-in-openai-competitor-anthropic.html}{Google}}). There also exists a rich body of open-source LLMs alongside these various closed-source offerings, often produced by these same companies. Though open-source only organizations, such as \textcolor{blue}{\href{https://nousresearch.com/}{Nous Research}}, also exist.

What follows is an analysis and explanation of UN Trade and Development's (UNCTAD) journey navigating this rapidly developing field, and how and why a completely open-source, custom LLM application was developed in-house. Section \ref{sec:rag} will explain RAG in more detail, section \ref{sec:opensource} will examine the open- and closed-source LLM landscapes, evaluating their pros and cons in the context of national statistics offices (NSOs) and international organizations. Section \ref{sec:unctadwork} will detail UNCTAD's work, experiences, and open-source libraries. Section \ref{sec:futurework} will examine future areas of improvement and work, while section \ref{sec:conclusion} will conclude.

\section{RAG}
\label{sec:rag}

\subsection{Why do we need RAG?} \label{section:whyrag}

An acute shortcoming of LLMs' architecture is their ignorance of information and events that lie outside of their initial training data. This means information that happens after their training or that wasn't initially included is inaccessible to the model. One solution would be to retrain the model daily or weekly and with text pertaining to your particular domain or use case. However, when \textcolor{blue}{\href{https://fortune.com/2024/04/04/ai-training-costs-how-much-is-too-much-openai-gpt-anthropic-microsoft/}{looking at the computing power and cost required to retrain an LLM from scratch}}, which can be in the millions of dollars, this quickly becomes infeasible. Though compute costs come down over time, LLM complexity and parameter count are set to continue to grow, meaning the cost of training frontier models will also continue to grow despite cheaper computing. This constrained knowledge base is a primary contributor to one of LLMs' most infamous characteristics, hallucinations. As the models work by predicting the next most likely token (or word, simplified) in a series, they can generate reasonable sounding responses even if they are not factually accurate. This is more likely to occur if a query is asked about information the model did not have in its training data.

\subsection{Context windows} \label{section:context_windows}

Now excluding training a model from scratch repeatedly due to cost and resource restrictions, what are solutions for making LLMs useful for more specific domains and current information and cutting down on hallucinations? The natural solution is to pass the new information alongside the prompt. LLMs have a concept window, which is the amount of tokens they are able to consider when generating a response. We can use this context window to pass information alongside our query. For instance, if we ask an LLM to "summarize the headline article from the BBC on this date", we may be greeted with a confident reply about one (more or less random) article of the likely many BBC articles contained in the training data. However, if we ask, "summarize this article, the full text of which is: ...", we will get a response grounded in the article we are actually interested in.

A limitation of this approach is the size of the context window, which varies by LLM. Context windows are expanding as the field develops, potentially to millions of tokens, but are still a limiting factor. The popular open-source LLama 2 LLM produced by Meta for instance has a context window of \textcolor{blue}{\href{https://www.ibm.com/topics/llama-2}{4,906 tokens}}. This is sufficient to pass the entirety of a news article to an LLM, but insufficient if the knowledge base extends to hundreds or thousands of domain-specific documents or datasets.

\subsection{Chunking and embedding} \label{section:chunking}

RAG addresses this problem by introducing an extra step into the generation pipeline. A document or large corpus of text is first split into smaller chunks, potentially enriched with metadata, then converted into an abstracted numerical vector, or embedding. Embedding models can capture semantic and contextual relationships between words and concepts. For instance, in the multi-dimensional vector space, "mountain" and "hill" would have embeddings that are close to each other, despite being far from each other in terms of letters.  

\subsection{Vector database and retriever} \label{section:vectordb}

These embeddings are then stored in a vector database. We now have our new information encoded in a searchable database. To make this information available to the LLM, we need to use a retriever to first determine the most relevant chunks of text to the query. The retriever embeds the query and searches for the N most relevant/closest text chunks in the vector database. The returned chunks are then passed alongside the query to the LLM to provide answers given the new context.

\subsection{Limitations} \label{section:limitations}

It is crucial to understand that the LLM is not simultaneously aware of all the information contained in the vector database, as is the case for information contained in its training data. Information in the training data is embedded directly into the LLM's parameters and weights. Rather, the base LLM is utilized for its knowledge base and what it has already learned about providing answers in natural language. The new text chunks retrieved from the vector database and passed alongside the query to the context is the only mechanism where the LLM is directly aware of new information. As a result, substantial components in the performance of a RAG LLM are the embedding model, vector database, and retriever, not just the LLM itself.

\section{Open-source vs. proprietary offerings}
\label{sec:opensource}

\subsection{Proprietary offerings} \label{section:proprietary}
A strong caveat to this section is the fact that the field is developing extremely rapidly, with new companies and products appearing and disappearing all the time. What is described here is relevant at the time of writing, but may no longer be in just a few months time. It is in no way meant to be a comprehensive overview of every product, library, and service available in the field, merely a quick overview for context. That being said, currently, when considering an organization's offerings when it comes to adopting a RAG LLM, paid options come in all sorts of shapes, sizes, and pricing models. One of the most "plug and play", user-ready RAG options is \textcolor{blue}{\href{https://blogs.microsoft.com/blog/2024/01/15/bringing-the-full-power-of-copilot-to-more-people-and-businesses/}{Microsoft's Copilot Pro}}, which costs 30 USD per user per month for businesses and can access documents stored in Sharepoint and query over them. Beyond that, there are a myriad of products, companies, APIs, and libraries for each component of a RAG LLM, including the embedding model (e.g,. \textcolor{blue}{\href{https://openai.com/blog/new-embedding-models-and-api-updates}{OpenAI}}), the vector database (e.g., \textcolor{blue}{\href{https://www.pinecone.io/}{Pinecone}}, \textcolor{blue}{\href{https://www.elastic.co/enterprise-search/vector-search}{Elasticsearch}}), and of course the LLM itself (e.g., \textcolor{blue}{\href{https://openai.com/pricing}{OpenAI}}, \textcolor{blue}{\href{https://docs.mistral.ai/platform/pricing/}{Mistral}}, \textcolor{blue}{\href{https://www.anthropic.com/api}{Claude}}). Implementing these types of solutions may still require coding. Smaller companies exist which may offer a more finished, user-friendly application while being built off of the APIs and services of the bigger players.

Building your own application but using paid components opens up the pipeline to potentially multiple cost producing touchpoints. For example, to run your application, you may be paying OpenAI for its embedding model, then paying Pinecone to store your embeddings in their vector database, then paying Anthropic to query against their Claude LLM, all while still needing to somehow manage the final user-facing interface.

\subsection{Open-source landscape} \label{section:opensource}
All of the functionality required to run a RAG LLM is also obtainable using exclusively open-source, free software, as will be detailed further in section \ref{sec:unctadwork}. A few of these libraries, frameworks, and resources will be mentioned here, but there exist many more. In terms of embedding models, there are dozens available for free on \textcolor{blue}{\href{https://huggingface.co/sentence-transformers}{Hugging Face}}, tuned for various general or specific use cases. \textcolor{blue}{\href{https://github.com/pgvector/pgvector}{Pgvector}} provides a Postgres-based vector database, \textcolor{blue}{\href{https://www.trychroma.com/}{Chroma DB}} offers another open-source vector database solution, while there are hundreds of available LLMs on \textcolor{blue}{\href{https://huggingface.co/models?other=LLM}{Hugging Face}}. Frameworks for creating RESTful APIs include \textcolor{blue}{\href{https://flask.palletsprojects.com/en/3.0.x/}{Flask}} and \textcolor{blue}{\href{https://fastapi.tiangolo.com/}{FastAPI}}, while options for front ends include \textcolor{blue}{\href{https://streamlit.io/}{Streamlit}} and \textcolor{blue}{\href{https://www.gradio.app/}{Gradio}}, among many others. For creating the final RAG pipeline, two notable libraries include \textcolor{blue}{\href{https://www.langchain.com/}{LangChain}} and \textcolor{blue}{\href{https://www.llamaindex.ai/}{LlamaIndex}}.

\subsection{Pros of in-house development} \label{section:pros}
\subsubsection{Cost} \label{section:pro_cost}
One of the simplest pros to developing generative AI applications in-house is cost. The cost of an in-house application comes down to the salary of the staff who initially develop it and the cost of the machine(s) where it is run. The latter can either be borne as an ongoing service fee by provisioning a cloud computer, or as a one-time fixed cost by buying a GPU(s). Here, it is important to note the uncertain pricing environment around managed solutions. In 2023, ChatGPT was estimated to spend \textcolor{blue}{\href{https://finance.yahoo.com/news/chatgpt-cost-bomb-openais-losses-125101043.html}{700,000 USD}} a day just operating ChatGPT. This figure also increases directly related to usage and the number of users, as the primary cost is compute, which is mutually exclusive between queries. As companies continue to race to gain a competitive advantage in the generative AI space, there is no proven, long-lasting, profitable, and sustainable established pricing model. That means that services may be offered at a loss simply to build market share, and may not be sustainable long term. This introduces the risk of becoming dependent upon a certain service for a RAG workflow, then being caught out if that service experiences a large cost increase. 

Per user pricing models are also common in the field. For example, to roll out Microsoft Copilot Pro to an organization of 500 people would cost 15,000 USD per month. To host the application UNCTAD has developed on a GPU-enabled Azure cloud machine supporting unlimited concurrent users costs about 144 USD per month, and can be made available to the entire organization. That cost is at on-demand pricing, it sinks to as low as 77 USD per month with longer-term agreements. Benefits come both in terms of lower operating costs, but also in terms of flexibility in terms of how and when those costs are incurred. 

The other, potentially larger difficulty than the cost itself in using paid solutions in such a fast-developing field is the institutional burden of getting new cost streams approved at large international organizations or NSOs. Waiting for a multi-month requisition process before being able to proceed in development is extremely difficult when the entire field is developing at lightning pace. Requested tools, services, or models may be outdated by the time resources are approved. By relying on free, open-source software, development is not constrained by resources or bureaucracy, but free to incorporate the latest developments as they are released.

\subsubsection{Flexibility} \label{section:pro_flexibility}

Flexibility in terms of features is another benefit of the in-house approach. Rather than receiving a finished product and relying on an external provider to add features at their discretion, the product is extendable and able to be improved on-demand. This results in greater flexibility in making the application relevant to how the organization wants to work with it. It also means the solution is not subject to change. A commercial offering is not guaranteed to continue working in its current form. The company may go bust, change their product offerings or pricing, or introduce censorship. With an in-house solution, it can persist in a given form for as long as desired, since every component is directly controlled. For instance, Google's Gemini has been updated to introduce stronger censorship, which has led to \textcolor{blue}{\href{https://new.pythonforengineers.com/blog/so-evidently/}{instances of over-correction}}. Google may fine-tune and improve its censorship in the future, but the key takeaway is the principle. A closed-source offering can be subject to change without your input, knowledge, or agreement, which can pose a problem if workflows are built around the assumption of a certain functionality being present or working as was originally intended.

\subsubsection{Fostering of institutional capacity} \label{section:pro_institutional_knowledge}

Benefits also include the accumulation and fostering of institutional knowledge. By researching and developing an in-house tool, the organization gains valuable know-how and the ability to work with generative AI not merely as a passive consumer, but also as an active developer. Many of the components developed for the RAG application have uses for other data products. For instance, the value of a vector database containing all of an organization's outputs that is searchable via natural language queries is not difficult to see, and is a direct byproduct of the RAG application development process. It also means the organization has a deep understanding of how the technology works, which is invaluable when it comes to training staff in the proper way to use and interpret model outputs. This also highlights the drawbacks of hiring temporary, external consultants to build a product. Delivery of a final, complete product is not particularly compatible with how the field has been developing. Better is to foster the capacities in-house, so the product can be continually developed and improved, rather than having a black box frozen in time which no one knows how to update. Additionally, in-house staff are able to repurpose components from what they have developed for different applications, something much more difficult to achieve with project-based consultants. Finally, as the organization gains a deeper understanding of generative AI's capabilities, limitations, development process, and potential applications, it can offer more comprehensive support and education in this area. This knowledge can inform UNCTAD's substantive work and expand the guidance and support it provides to developing countries on the topic.

\subsubsection{Privacy and data control} \label{section:pro_privacy}

A last benefit is privacy and data control. By developing an in-house system, organizations can be sure that their data is stored only where they want it to be. This includes not only the actual documents provided to the RAG system, but also the queries that staff members send. The former is important when you are working with confidential or private data. The latter is an often overlooked aspect of LLMs, where companies like OpenAI \textcolor{blue}{\href{https://www.androidauthority.com/does-chatgpt-save-data-conversations-3310883/}{store users' chat history}}. ChatGPT and other services are free not only to acquire new users and increase usage of generative AI in general and thus grow the market for their services, but also to train future versions of the model. This opens up personal and work conversations to not only those companies, but potentially the entire user base of future versions, and potentially advertisers. Though OpenAI says they do not currently sell chat history data, nothing is stopping them from doing so in the future. To say nothing of the potential for data breaches. None of which is a risk with an in-house solution, as the organization maintains full control of what is saved, whether it is saved, and where it is saved.

\subsection{Cons of in-house development} \label{section:cons}

\subsubsection{Availabilty of resources and capacity} \label{section:con_capacity}

There are of course substantial cons associated with the in-house approach. The key one is availability of staff with the proper skill set to set up and develop such a system. Though international organizations and NSOs are focusing increasingly on hiring data scientists and data engineers with the skills to actualize such projects, there remains a sizable skills gap.

\subsubsection{Scalability} \label{section:con_scalability}

Another drawbacks includes scalability. By going it on your own, you are responsible for handling scaling of the application. What may work fine for 10 users may be more difficult to get working for 100 or 10,000 users. Paid solutions have hundreds of people working on infrastructure and scaling, meaning this issue is abstracted away from application development. 

\subsubsection{Polish} \label{section:con_polish}

Polish, robustness, and feature set are other important aspects. Companies with hundreds or thousands of employees will naturally be able to make much more robust and polished user interfaces than can be made internally. This can include things like aesthetics and design, robustness to different edge cases and system states, and the addition of more quality of life and substantive features. These are all easier to achieve with dedicated staff developing a single service for hundreds or thousands of paying customers.

\subsubsection{Access to frontier models} \label{section:con_frontier}

Finally, there is the issue of access to frontier models. Most of the state-of-the-art models in terms of power and performance are closed-source. Open-source models are likely to always remain a step behind.

While I have presented the options of an open-source, in-house solution and a paid solution as a binary, the reality is much more nuanced. There is nothing stopping you from using open-source for your front end and vector database, while using OpenAI's LLM API for the "brains" of your model, or vice versa. From one end to the other is a spectrum that will likely differ from organization to organization based on their resources and needs.

\section{UNCTAD's work and experience}
\label{sec:unctadwork}

\subsection{Technical details} \label{section:subst}

This section will detail the work UNCTAD has done so far in developing its in-house application. Much like the field itself, work is ongoing and likely to change and expand after the time of writing, so check the mentioned repositories for the latest developments.

The first component necessary in a RAG application is the processing of the text corpus into a format that works with the embedding model. UNCTAD had already previously developed a \textcolor{blue}{\href{https://github.com/dhopp1/nlp_pipeline}{library}} for the processing of PDF (including optical character recognition conversion for scanned documents), Word, and HTML documents into raw text. While LlamaIndex provides document loaders for these file types as well, the library has the added benefit of easily adding metadata information to the texts, as well as enabling statistical analysis and topic modelling.

LlamaIndex was the primary tool used for the RAG LLM. The vector database is stored in a local PostgreSQL and made compatible with embeddings and vector search via \textcolor{blue}{\href{https://github.com/pgvector/pgvector}{pgvector}}. The final pipeline was packaged together in another \textcolor{blue}{\href{https://github.com/dhopp1/local_rag_llm}{library}}. The library is LLM-agnostic, accepting any LLM stored in GGUF format, as well as embedding model-agnostic.

The \textcolor{blue}{\href{https://github.com/dhopp1/streamlit_rag}{final user-facing front end}} was written in \textcolor{blue}{\href{https://streamlit.io/}{Streamlit}} and includes provisions for multiple simultaneous users and persistence of vector databases. The latter is important so that large corpora do not need to be revectorized every use.

\begin{figure}[htp]
    \centering
    \includegraphics[width=13cm]{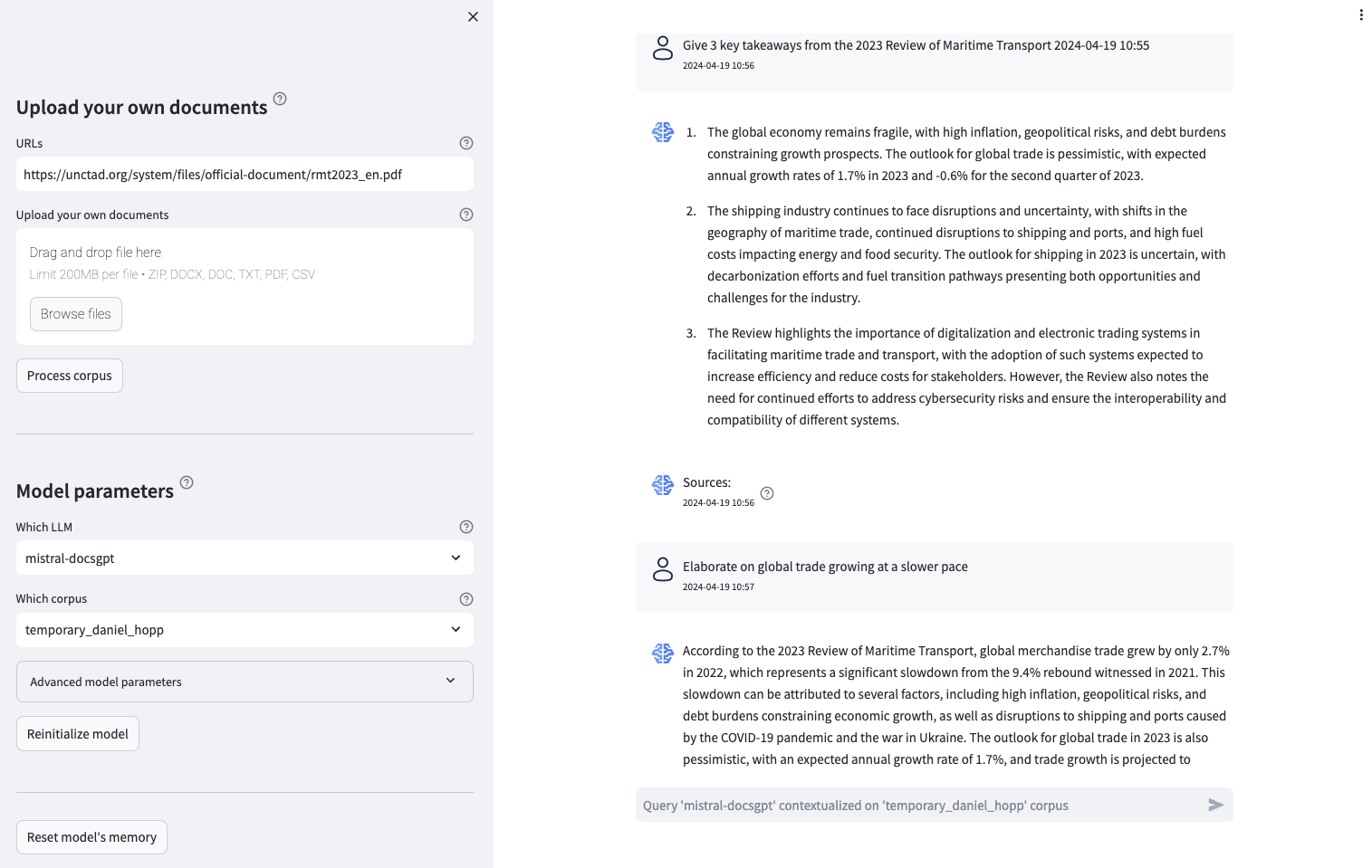}
    \caption{Example screen of the application}
    \label{fig:example}
\end{figure}

All three libraries are available on GitHub and PyPI and can be installed to duplicate the application on any machine. Dockerfiles are also available to run the application, further lowering the barrier to entry. Depending on the size of the LLM you use, it runs well on any Nvidia GPU-enabled machine, as well as on Apple silicon Macs. It can also run on any computer's CPU, albeit more slowly, though still usable. Unlimited simultaneous user sessions are supported. The only constraint is simultaneous generation, where requests received at the same time are queued and executed in the order received. Figure \ref{fig:example} shows an example image of the application.

\subsection{Institutional space and support} \label{section:space}

One of the main findings in the course of the development of the application was the importance of institutional space for pursuing innovative or long shot data products. Data science projects are difficult to perform under the burden of traditional work programs and mandate structures. This and other similar ones (such as the \textcolor{blue}{\href{https://github.com/dhopp1/nlp_pipeline}{NLP pipeline library}}) were developed without a clear application or request in mind. Despite this lack of an immediate recognized need, they have gone on to have the potential for transformative impact across the organization's work. Further, developed and maintained correctly, they tend to build and compound in utility over time. For instance, besides the aforementioned natural language search vector database, the front-end interface can be repurposed for other internal or external data products. Work of this type is more akin to building a platform to enhance other work than producing an output in and of itself. 

All of the described work was carried out by UNCTAD's Global Crisis Response Group (GCRG). These and other innovations, not only around data products but also around working methods, communication formats, and research approaches, were enabled by the GCRG's unique (within UNCTAD) institutional setting. With no legacy reports or work programs to fulfill, the group could focus on quick-response innovation projects. With this dedicated space, innovation could bloom, not rely on staff working nights and weekends on passion projects alongside their regular work, or on slow and incremental steps, or, as is often the case, not happen at all.

As such, it is my hope that successful projects like this serve to further convince the official statistics and international organization community of the importance of investing in not only data science capacities, but institutional space for the field.

\section{Future work}
\label{sec:futurework}

\subsection{Fine-tuning} \label{section:finetuning}
As with any data product, there are still many avenues for improvement. One is the application of fine-tuning to an existing LLM model. Fine-tuning involves freezing the majority of the weights of a pre-trained model, and only training a small subset of layers for the task or domain of interest. Since only a small subset of weights are adjusted, the computational and cost requirements of fine-tuning are substantially less than that of pre-training from scratch. This enables the LLM to maintain the knowledge base, sentence construction abilities, etc., that it learned from pre-training, while making it more suitable for a particular task. Fine-tuning can be used for either knowledge injection or task augmentation. For instance, if a base model is originally trained for text completion, as is usually initially the case, it won't be particularly useful for chat interfaces. Fine-tuning it on question-response pairs then enables it to perform much better with prompts in this style. While fine-tuning has been found to be \textcolor{blue}{\href{https://arxiv.org/abs/2312.05934}{less useful for specific fact retrieval}} than RAG in many instances, it has the benefit of making the LLM simultaneously aware of a large corpus of information, rather than relying on the embedding model and vector search function and being constrained by the context window in terms of how much information the actual LLM itself can pull from when responding to queries. The best solution likely combines both approaches, using RAG on a fine-tuned LLM. For instance, we could fine-tune an LLM on the whole corpus of UNCTAD publications, specializing it on UNCTAD's positions, style, and body of work, then use RAG for direct information retrieval varying by the specific application. 

Alongside the other three libraries, the \textcolor{blue}{\href{https://github.com/dhopp1/local_llm_finetune}{local\_llm\_finetune}} library was also developed. The library, usable via Docker, leverages the open-source \textcolor{blue}{\href{https://unsloth.ai/}{Unsloth}} package to enable local fine-tuning on corpora of raw text, without question and answer pairs, for knowledge and style injection.

\subsection{RAG optimization} \label{section:rag2}
Another avenue of improvement involves improving the existing RAG system. This includes things like fine-tuning or improving the embedding model and vector search functions in isolation. This would ensure that the most relevant and helpful text chunks are being passed to the LLM for answers. Hyperparameters can also be optimized, such as researching which chunk sizes, chunk overlap values, context window sizes, etc., produce the most helpful answers for the different use cases UNCTAD has in mind. Introducing pre-filtering steps to the vector database may also be helpful. For instance, rather than relying on the retriever to parse from your query that you are only interested in documents from a certain year, the database could be pre-filtered for that year before the search, dramatically shrinking what needs to be searched and increasing the likelihood of returning relevant and useful chunks for the LLM.

\subsection{Rollout, adoption, and training} \label{section:rollout}
Final next steps include the process of rollout, adoption, and training. These steps are crucial for any new data product, but particularly for LLMs. Because of their natural language responses, they can appear omniscient and confident in any response they give. If users do not understand how they work and their limitations and do not follow correct validation guidelines, this can lead to hallucinations and wrong information making its way into official reports or other pieces. RAG in particular helps with this issue, as the application returns alongside its answers both the actual texts it is basing its answer off in addition to the metadata for those texts. So it is relatively easy to verify that if a fact or figure the LLM returns is not contained in the source texts, it is likely a hallucination. Still, this requires training and awareness for end users.

\section{Conclusion}
\label{sec:conclusion}
Generative AI offers much promise for NSOs and international organizations, just as it does for the private sector. While it is wholly possible they will benefit from it by being bystanders, waiting for large tech companies to deliver ready-made solutions on their terms, the field's rich open-source community and resources means they are also able to exercise agency in this new era. UNCTAD's experience has shown that with the proper investment in data science capacity and institutional space, it is possible to develop generative AI applications entirely for free and in-house, customized to an organization's specific needs. Additionally, developing solutions on your own or leveraging paid services is not a binary. The developed infrastructure and institutional knowledge acquired in this process do not preclude UNCTAD from introducing paid components into the pipeline, or switching to a completely managed solution. For instance, in the future the LLM could be changed from local open-source to an OpenAI API call, releasing the requirement for hosting on a GPU-equipped machine, addressing scale, and giving access to frontier models all while continuing to not incur costs for the vector database or embedding model. Development of in-house skills increases flexibility. Hopefully, by sharing UNCTAD's experience and contributing to the vibrant open-source community around generative AI with the tools it has developed, we can lower the barrier to entry for NSOs and international organizations in developing their own solutions. 

% Bibliography.
%\bibliography{references}

\end{document}